\documentclass{article}
\usepackage{spconf,amsmath,graphicx,hyperref}
\usepackage{booktabs}
\usepackage{subcaption}
\usepackage{amssymb,amsfonts}
\usepackage{multirow} 

\title{VividVoice: A Unified Framework for Scene-Aware Visually-Driven Speech Synthesis}
\name{Chengyuan Ma$^1$, Jiawei Jin$^1$, Ruijie Xiong$^2$, Chunxiang Jin$^2$, Canxiang Yan$^2$, Wenming Yang$^{1*}$ \thanks{$^*$ Corresponding authors}
\thanks{This work was supported in part by the National Key R\&D Program of China (2023YFB4302200) and the Special Foundations for the Development of Strategic Emerging Industries of Shenzhen (KJZD20231023094700001).}}
\address{$^1$ Shenzhen International Graduate School, Tsinghua University, China,
$^2$Ant Group, China
}
\begin{document}
\ninept
\maketitle
\begin{abstract}
We introduce and define a novel task—Scene-Aware Visually-Driven Speech Synthesis, aimed at addressing the limitations of existing speech generation models in creating immersive auditory experiences that align with the real physical world. To tackle the two core challenges of data scarcity and modality decoupling, we propose VividVoice, a unified generative framework. First, we constructed a large-scale, high-quality hybrid multimodal dataset, Vivid-210K, which, through an innovative programmatic pipeline, establishes a strong correlation between visual scenes, speaker identity, and audio for the first time. Second, we designed a core alignment module, D-MSVA, which leverages a decoupled memory bank architecture and a cross-modal hybrid supervision strategy to achieve fine-grained alignment from visual scenes to timbre and environmental acoustic features. Both subjective and objective experimental results provide strong evidence that VividVoice significantly outperforms existing baseline models in terms of audio fidelity, content clarity, and multimodal consistency. Our demo is available at  \url{https://chengyuann.github.io/VividVoice/}
.
\end{abstract}
\section{Introduction}








\vspace{-2mm}

Scene-Aware Immersive Speech Generation aims to achieve the joint synthesis of two essential acoustic components: (1) speech with controllable content, and (2) environmental acoustics that align with a specific scene. This task goes beyond the traditional paradigm of speech synthesis, which primarily focuses on generating clean speech, by expanding the expressive dimension to simultaneously convey linguistic information and spatial environmental context \cite{lee2024voiceldm,im2024diffrent,jung2025voicedit,vyas2023audiobox}. Achieving such synchronized generation of speech and environment is crucial for enhancing perceptual realism and user immersion in next-generation applications such as virtual reality and digital humans \cite{ahmed2024voick,kurniawati2012personalized}.


Early approaches to environment-aware speech generation primarily simulated static acoustic characteristics by overlaying clean speech with white noise or fixed reverberation. To pursue more realistic dynamic effects, subsequent research shifted toward acoustic feature transfer, such as extending environmental control by extracting room impulse responses (RIRs) \cite{tan22_interspeech_RIR}, or transferring noise and reverberation features from reference audio to target speech, as in DiffRENT \cite{im2024diffrent} and AST-LDM \cite{kim24e_interspeech_astldm}. Although these methods improved realism, their controllability remains constrained by the availability of reference signals, making it difficult to flexibly generate speech for arbitrary new scenes. To enable more flexible generative control, studies represented by VoiceLDM \cite{lee2024voiceldm} adopted a text-driven approach, attempting to jointly model speech and environmental sounds through natural language descriptions. Nevertheless, the text modality has an inherent limitation in information density: its abstractness and ambiguity make it insufficient to capture the rich acoustic details embedded in visual scenes. Thus, turning to the visual modality, which is denser and more detail-rich, becomes a natural choice for overcoming this limitation.


However, current vision-driven studies fall short of meeting the requirements of our task. One line of work (e.g., FaceTTS \cite{goto2020face2speech}) focuses on the speaker, generating timbre consistent with visual identity but lacking the ability to perceive or model environmental acoustics. Another line of work (e.g., SSV2A \cite{guo2024gotta_ssv2a}) emphasizes the scene, producing realistic environmental soundscapes but failing to synthesize specified linguistic content. This inability to jointly generate speech and environmental acoustics arises from two fundamental challenges at both the data and model design levels. First, on the data level, a strongly aligned dataset that simultaneously includes visual scenes, speaker identities, and corresponding environmental acoustics is still missing. For example, AudioSet \cite{audioset1}provides audiovisual scenes but lacks stable speaker identity, while datasets such as LRS3-TED \cite{afouras2018lrs3} contain aligned speech and identity but are typically recorded in acoustically “clean” conditions, preventing models from learning correlations between scenes and acoustics. Second, on the model design level, existing alignment architectures are insufficient for learning the complex many-to-one mapping from a single visual input to the decoupled components of timbre and environmental acoustics. As a result, they struggle to preserve fine-grained features while achieving independent control of both attributes.





To address the aforementioned limitations, we propose a unified solution centered on a new task—Scene-Aware Visually-Driven Speech Synthesis. Our main contributions are as follows:

(1) We pioneeringly propose the novel task of Scene-Aware Visually-Driven Speech Synthesis and design its first unified generative framework, VividVoice, which aims to generate immersive speech that is acoustically well-coupled with the visual scene.

(2) To support this task, we have constructed the first large-scale dataset, Vivid-210K. This dataset, built through an innovative programmatic pipeline, provides the strong multimodal correlation information required for training.

(3) We propose an acoustic feature alignment method based on decoupling. The core of this method is our designed Decoupled  Multimodal Scene-Voice Alignment (D-MSVA) module, which is trained using a unique hybrid supervision strategy to achieve independent control over speaker timbre and environmental acoustics.

\vspace{-2mm}
\vspace{-2mm}
\section{Method}
\vspace{-2mm}
\subsection{Vivid-210k}



\begin{figure}[htb]
\label{main_fig}
  \centering
  \includegraphics[width=0.9\linewidth]{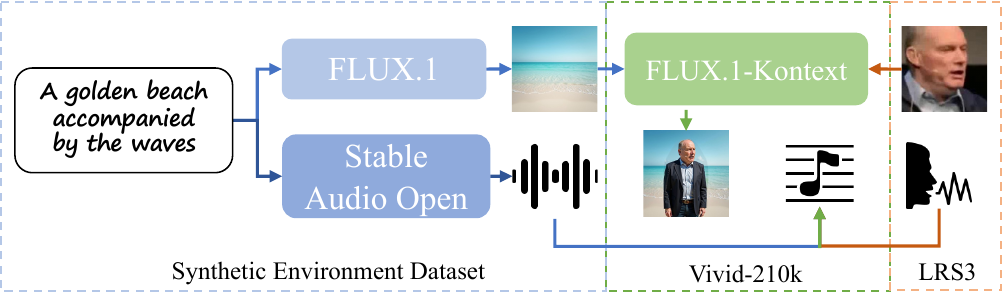}
  \vspace{-3mm}
  \caption{The programmatic pipeline for constructing our Vivid-210K dataset.}
  \label{fig:data}
  \vspace{-4mm}
\end{figure}
  \vspace{-1mm}

To support the proposed task of Scene-Aware Visually-Driven Speech Synthesis, we constructed a large-scale multimodal dataset, Vivid-210K, which contains over 210k samples and more than 800 speakers. The dataset consists of a procedurally synthesized pre-training set and a real-world fine-tuning set designed to narrow the domain gap. Our innovative procedural synthesis pipeline (as illustrated in Fig. \ref{fig:data}) follows the data flow described below: first, we extract speaker identities (facial images and clean speech) from the LRS3 dataset. To obtain matching audiovisual environments, we examined existing datasets such as VGGSound, but found that they generally suffer from limitations in both quality and alignment. Therefore, we adopt a “paired generation” strategy—using the same textual prompt to drive both a text-to-image model (FLUX.1) and a text-to-audio model (Stable Audio Open \cite{evans2025stable})—to ensure semantic consistency of the generated scenes. Subsequently, during the data fusion stage, we employ the instruction-guided image editing technique FLUX.1-Kontext-dev \cite{batifol2025flux} to seamlessly integrate facial images into the scene, and mix the corresponding speech with environmental sounds under a random signal-to-noise ratio. For the fine-tuning set, we extract and decompose independent visual backgrounds, environmental sounds, and speech audio elements from real videos to improve the model’s generalization capability. 

To ensure dataset quality and cross-modal consistency, we design an automated evaluation pipeline based on collaboration between vision–language models (VLMs) and large language models (LLMs). Specifically, Qwen2.5VL \cite{bai2025qwen2.5} generates structured descriptions of synthetic images, which are then jointly evaluated by a panel of LLMs (Qwen3 \cite{yang2025qwen3}, DeepSeek-V3 \cite{liu2024deepseek}, and GPT-4o \cite{hurst2024gpt}) to score the semantic alignment between image content and scene labels, filtering out low-quality samples. Through this process, the automatic scene-matching rate of synthetic samples reached 98.6\%. Furthermore, subjective evaluation conducted by domain experts showed an audiovisual consistency rate as high as 95.4\%. These results strongly demonstrate the reliability of the Vivid-210K dataset, providing a solid foundation for model training.

\vspace{-2mm}
\subsection{System Overview}
\begin{figure*}[t]
\label{main_fig}
  \centering
  \includegraphics[width=0.85\linewidth]{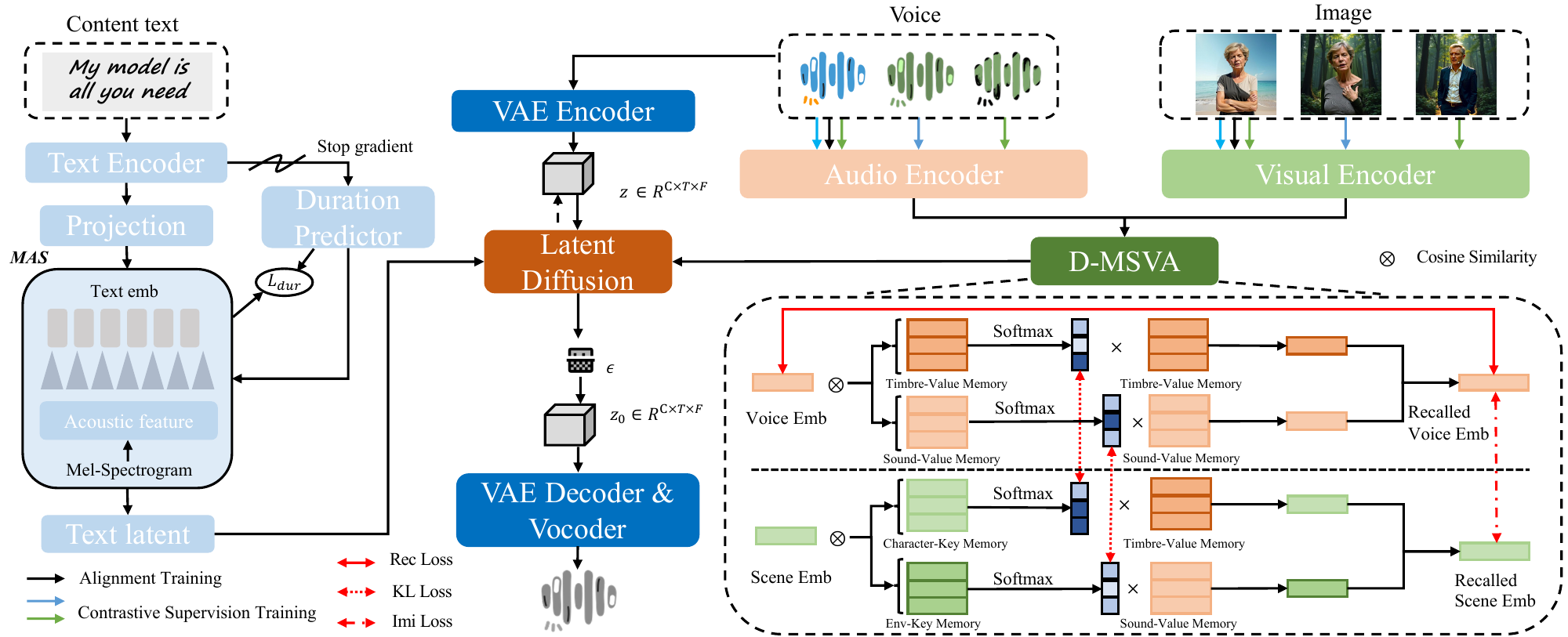}
  \vspace{-3mm}
  \caption{
The overall architecture of the VividVoice framework.} 
  \label{fig:framework}
  \vspace{-5mm}
\end{figure*}

The proposed VividVoice framework is designed to achieve end-to-end visually-driven scene-to-speech synthesis. As illustrated in Fig. \ref{fig:framework}, the overall architecture consists of three key components: a Content Generation Pathway,  a Scene Perception Pathway, and a Latent Diffusion Backbone. In the Content Generation Pathway, we follow the successful paradigm established by VITS \cite{kim2021vits}. The input text is first processed by a Text Encoder and a Duration Predictor, and aligned with acoustic feature frames using Monotonic Alignment Search (MAS), producing text latents that are temporally aligned with the acoustic representation. This pathway guarantees that the synthesized speech maintains high accuracy in linguistic content, along with natural prosody and rhythm. In the Scene Perception Pathway, the model receives multimodal scene inputs, including speech (Voice) and images (Image). To extract semantic representations, we employ the pretrained CLAP model as the Audio Encoder and the pretrained MetaCLIP \cite{xu2023metaclip} model as the Visual Encoder. These representations are then passed into our core module, D-MSVA, which performs fine-grained alignment and fusion of cross-modal scene information, ultimately producing a unified Recalled Scene Embedding. Finally, the text latents from the Content Generation Pathway and the Recalled Scene Embedding from the Scene Perception Pathway jointly condition a Latent Diffusion Model, which iteratively performs denoising to generate acoustic latents $z_0$. These latents are then decoded through a pretrained VAE Decoder followed by a vocoder, yielding high-fidelity speech waveforms that are acoustically well-matched to the visual scene.

\vspace{-2mm}

\subsection{Decoupled Memory-based Scene-Voice Alignment}




Inspired by MFVA \cite{sheng2023facedriven}, we propose a novel Decoupled Multi-modal Scene–Voice Alignment (D-MSVA) module to achieve fine-grained alignment between complex visual scene information and mixed audio signals. Unlike conventional alignment modules, the core idea of D-MSVA is to decouple highly entangled multi-modal features into two independent and semantically meaningful components—speaker timbre and scene sound effects—at the architectural level through multiple specialized memory banks.

The module is composed of four learnable memory banks ($M_{\{\cdot\}} \in \mathbb{R}^{N \times D}$), which jointly establish a bridge from visual space to auditory space: the Character-Key Memory Bank ($M_{pk}$) and the Environment-Key Memory Bank ($M_{ek}$) for encoding visual concepts, together with the Timbre-Value Memory Bank ($M_{tv}$) and the Sound-Value Memory Bank ($M_{sv}$) for storing auditory primitives. Here, $N$ represents the number of slots in each decoupled concept memory bank (i.e., character/timbre and environment/sound), and $D$ denotes the embedding dimension. During training, D-MSVA is optimized through a parallel dual-path mechanism. The auditory pathway is designed to ensure the representational capacity of the auditory memory banks. Given a mixed auditory embedding $a \in \mathbb{R}^D$ that contains both timbre and sound effects, this pathway queries the value memory banks ($M_{tv}$ and $M_{sv}$) to disentangle its components, computing two sets of reference attention weights, $w'_{t}$ and $w'_{s}$. For example, the timbre weight $w'_{t,i}$ is computed as:

\begin{equation}
w_{t,i}^{\prime}=\mathrm{softmax}\left(\frac{a^Tm_{tv,i}}{||a||_2||m_{tv,i}||_2}\right)
\end{equation}

where $m_{tv,i}$ represents the $i$-th learnable slot vector in the Timbre-Value Memory Bank. The computation of the sound weights $w'_{s}$ follows in an analogous manner. Using these two weight sets, the timbre component and the sound-effect component are reconstructed from their corresponding value memory banks and combined into the final reconstructed auditory embedding $\hat{a}$. This process is supervised by minimizing the auditory reconstruction loss $L_{rec}$. Meanwhile, in a manner similar to Eq. (1), given a mixed visual embedding $v \in \mathbb{R}^D$ that contains both character and environmental information, the visual pathway queries the key memory banks to compute the corresponding character weights $w_p$ and environment weights $w_e$. After obtaining the two sets of disentangled visual attention weights, the core task of this pathway is cross-modal retrieval and synthesis: $w_p$ is used to query the Timbre-Value Memory Bank ($M_{tv}$), and $w_e$ is used to query the Sound-Value Memory Bank ($M_{sv}$). The results are then combined to form the “recalled” scene embedding:
$\hat{a}_v = M_{tv}^T w_p + M_{sv}^T w_e$. To enforce cooperation between the two pathways, we introduce an additional Attention Alignment Loss $L_{align}$, which regularizes the consistency between the attention weight distributions of the auditory and visual pathways via KL divergence:
\begin{equation}
    L_{align} = D_{KL}(w'_{t} || w_p) + D_{KL}(w'_{s} || w_e)
\end{equation}
Through the joint supervision of reconstruction and alignment, the D-MSVA module is trained end-to-end to learn a fine-grained, decoupled mapping from visual scenes to their corresponding acoustic features.

\vspace{-2mm}

\subsection{Inter-Modal Mixed Supervision Strategy for Decoupling}


To fully exploit the decoupling potential of the D-MSVA module, we design a hybrid supervision strategy. This strategy is divided into Alignment Supervision and Contrastive Disentanglement Supervision. When the training samples consist of standard paired data , we apply Alignment Supervision, which aims to learn an accurate mapping between the modalities. This component of the loss consists of three parts: auditory reconstruction loss ($L_{rec}$), attention alignment loss ($L_{align}$), and imitation loss ($L_{imi}$). The imitation loss ensures that the decoupled components generated by the visual pathway can precisely mimic the corresponding components decoupled by the auditory pathway:
\begin{equation}
L_{imi} = ||a'_{timbre} - \hat{a}_{timbre}||_2^2 + ||a'_{sound} - \hat{a}_{sound}||_2^2
\end{equation}
When we sample data pairs with specific attribute differences from the dataset, we introduce Contrastive Disentanglement Supervision to explicitly guide the model in learning decoupling. Specifically, when given data pairs from the same character but in different environments, we impose the Timbre Consistency Loss ($L_{timbre\_c}$) to ensure that the timbre features generated by the visual input of the same character remain consistent:
\begin{equation}
L_{timbre\_c} = ||\hat{a}_{timbre\_A} - \hat{a}_{timbre\_B}||_2^2
\end{equation}
On the other hand, when given data pairs from different characters but in the same environment, we impose the Environment Consistency Loss ($L_{env\_c}$) to ensure that the environmental sound features generated by visual input of the same environment are as similar as possible:
\begin{equation}
L_{env\_c} = ||\hat{a}_{sound\_A} - \hat{a}_{sound\_B}||_2^2
\end{equation}
In summary, the total loss function for the D-MSVA module during training is a weighted sum of all the loss components as follows:
\begin{equation}
L = L_{rec} + \lambda_{1}L_{align} + \lambda_{2}L_{imi} + \lambda_{3}L_{timbre\_c} + \lambda_{4}L_{env\_c}
\end{equation}
Where $\lambda_{1}$, $\lambda_{2}$, $\lambda_{3}$, and $\lambda_{4}$ are constant weights that control the importance of each component.
\section{Experiments}
\begin{table*}[htb]
\centering
\caption{
The experimental results of the benchmark evaluation for different methods. Here, $\uparrow$ indicates that higher values are better, while $\downarrow$ indicates that lower values are better.}
\vspace{-3mm}
\resizebox{0.8\linewidth}{!}{
\begin{tabular}{l|cccc|cccc}
\toprule
Model & WER(\%)$\downarrow$&FAD$\downarrow$      & KL$\downarrow$&CLAP$_{cap}$ $\uparrow$ &   MOS-CO$\uparrow$ & MOS-TI $\uparrow$ & MOS-SC $\uparrow$& MOS-NA $\uparrow$ \\ \midrule
GT & 10.62           & -     & -    &  0.39  &  4.36   & 4.03 & 4.11 & 4.25 \\ \midrule
VoiceLDM & 9.23  &   4.74   &   1.79  & \bf{0.27}   & 3.23    &  1.75 & 2.56 & 3.41 \\

VividVoice(Ours)& \bf{7.15}  & \bf{3.98}   & \bf{1.53}   & 0.25     & \bf{3.95}   & \bf{3.08}   &  \bf{4.30}     & \bf{3.88}   \\ 
\bottomrule
\end{tabular}}
\label{table:main1}
\vspace{-5mm}
\end{table*}

\subsection{Implementation Details}

All our experiments are based on the proposed Vivid-210K Dataset. We use 160k samples for training, with the remaining samples reserved for validation, and 12 unseen speakers from the real-world fine-tuning set as the test set to evaluate performance in realistic scenarios. All audio samples are downsampled to 16kHz, and when audio is fused with environmental sound effects, a random signal-to-noise ratio (SNR) between 4 and 20 dB is applied. We adopt the same U-Net structure as the AudioLDM \cite{liu2023audioldm} diffusion model and train it from scratch. The main training phase is conducted over 800,000 steps using 8 NVIDIA A100 GPUs, with a batch size of 8 per GPU. The AdamW optimizer is employed with a fixed learning rate of 1e-5. For the D-MSVA module, the number of slots in the internal memory banks is set to $N=128$, and the weight hyperparameters for the individual loss components are set as follows: $\lambda_{1}=10$, $\lambda_{2}=2$, $\lambda_{3}=0.5$ and $\lambda_{4}=0.5$. The subsequent fine-tuning phase lasts for 100,000 steps, using exponential moving average (EMA) \cite{klinker2011ema} and automatic mixed precision (AMP) \cite{micikevicius2017amp} processing, with an effective batch size of 16 and a learning rate adjusted to 5e-5. 
\vspace{-3mm}
\subsection{Metrics}
\vspace{-2mm}


\noindent \textbf{Objective Metrics:} 
To quantitatively evaluate the performance of the VividVoice model, we use a set of objective metrics. Word Error Rate (WER) is used to measure content accuracy and clarity (calculated by a pretrained Whisper-Large-v3 model). The overall audio fidelity is evaluated through Fréchet Audio Distance (FAD)  and KL Divergence (KL) , which compare the similarity of feature distributions and probability distributions between the generated and real audio, respectively. To assess the core audiovisual consistency, we use the CLAP$_{cap}$ score \cite{wu2023largeclap}, which measures the quality of the visual-to-acoustic conversion by calculating the matching score between the generated audio and the original image caption.

\noindent \textbf{Subjective Metrics:}
For subjective evaluation, we conducted a crowdsourced listening test where 25 evaluators provided Mean Opinion Scores (MOS) on a scale of 1 to 5 across four dimensions. The four dimensions are: Overall Naturalness (MOS-NA), which evaluates the overall perceptual quality of the audio; Scene Consistency (MOS-SC), which assesses the degree of alignment between the acoustic environment of the audio and the visual scene; Timbre Consistency (MOS-TI), which evaluates the degree to which the timbre characteristics of the generated speech align with the character defined by the image; and Content Fidelity (MOS-CO), which measures the intelligibility and content accuracy of the speech.

\vspace{-3mm}
\subsection{Evaluation Results}
\vspace{-2mm}

Since existing vision-driven models typically address only single-attribute mappings without linguistic content, they were excluded from quantitative comparison due to task mismatch. Therefore, we select VoiceLDM as the primary baseline for comparison, as its ultimate goal—generating immersive speech with environmental scene awareness—is the closest to our work. As shown in Table \ref{table:main1}, our model significantly outperforms the baseline on the real-world test set across the majority of both objective and subjective evaluation metrics. In terms of speech clarity and audio fidelity, VividVoice achieved a 7.15\% Word Error Rate (WER), which is not only lower than VoiceLDM’s 9.23\%, but also better than the real recordings (GT). Furthermore, it achieved superior scores in both FAD  and KL, with values of 3.98 and 1.53, respectively, demonstrating its exceptional content retention capability and higher audio generation quality. Subjective evaluations reveal that our model scored 3.08 and 4.30 in Timbre Consistency (MOS-TI) and Scene Consistency (MOS-SC), respectively, far exceeding VoiceLDM's scores of 1.75 and 2.56. Although the CLAP$_{cap}$ score is slightly lower due to interference from non-sounding visual details in captions, this metric bias does not reflect audio fidelity and could be mitigated by relevance filtering in future work. These results strongly demonstrate the superiority of the VividVoice framework in overall performance, achieving both high-fidelity audio generation and high consistency with visual scenes, thus providing a more immersive auditory experience.
\vspace{-2mm}
\subsection{Ablation Study}
\vspace{-2mm}

\begin{figure}[htb]
\label{main_fig}
  \centering
  \includegraphics[width=0.7\linewidth]{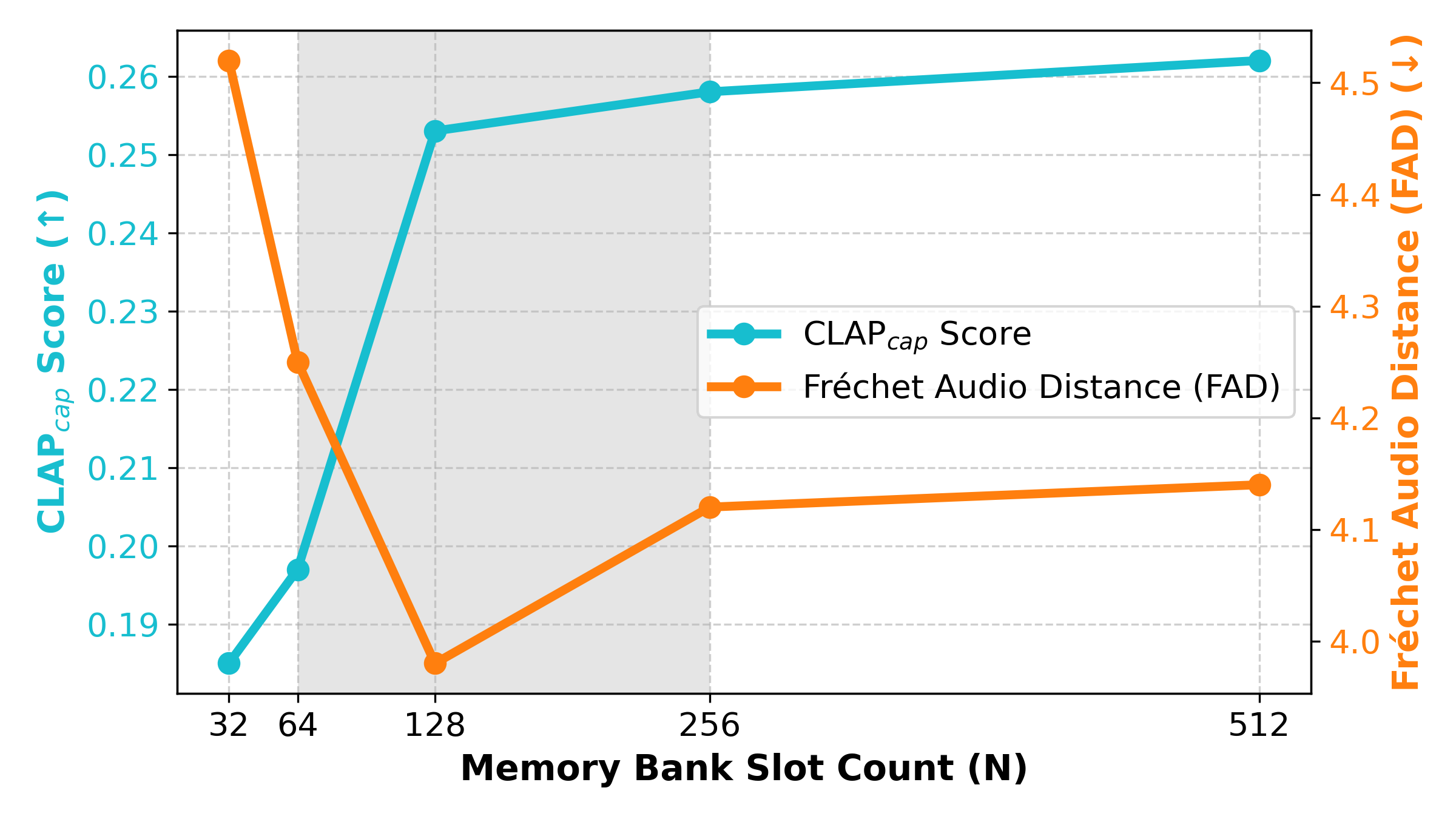}
  \vspace{-3mm}
  \caption{
Ablation study on the number of memory bank slots.} 
  \label{fig:SLOTS}
  \vspace{-4mm}
\end{figure}

To validate the effectiveness of the proposed D-MSVA module in multimodal alignment and feature decoupling, we conducted an ablation study. As shown in Table \ref{table:main2}, we compared the complete VividVoice model with two mainstream fusion methods: (1) w/ Concat-Fusion, a simple baseline that directly concatenates multimodal embeddings; and (2) w/ Attn-Fusion, a strong baseline that uses a standard cross-attention mechanism \cite{vaswani2017attention} for fusion. The experimental results show that our proposed D-MSVA achieves the best performance across all objective metrics. Specifically, compared to the stronger Attn-Fusion baseline, our method achieved a 16.0\% and 9.5\% relative improvement in FAD and KL, respectively. This result strongly demonstrates that D-MSVA, through its unique decoupled memory bank design and hybrid supervision strategy, can more effectively fuse multimodal information and convert it into high-quality audio generation. In contrast, simple concatenation or attention mechanisms struggle to achieve the same level of fine-grained alignment and control. We further investigate the impact of the number of memory bank slots ($N$) in the D-MSVA module, a parameter that jointly determines the representational capacity for the two decoupled concepts: “character/timbre” and “environment/sound.” As shown in Fig. \ref{fig:SLOTS}, while a larger $N$ improves alignment accuracy (CLAP$_{cap}$), the audio generation quality (FAD) begins to degrade due to overfitting when 
$N$ exceeds 128. Therefore, we recommend setting the value of $N$ to 128.

\vspace{-2mm}
\begin{table}[htb]
\centering
\caption{
Ablation experiment results for the D-MSVA module. w/ Attn-Fusion uses the standard attention-based fusion module, while w/ Concat-Fusion uses a simple concatenation fusion module.}
\vspace{-3mm}
\resizebox{0.9\linewidth}{!}{
\begin{tabular}{l|cccc}
\toprule
Model & WER(\%)$\downarrow$&FAD$\downarrow$      & KL$\downarrow$&CLAP$_{cap}$ $\uparrow$ \\ \midrule
VividVoice w/ Concat-Fusion& 7.85          & 4.87     & 1.85    &  0.18   \\ 
VividVoice w/ Attn-Fusion & 7.31  &   4.74   &   1.69  & 0.24    \\

VividVoice(D-MSVA)& \bf{7.15}  & \bf{3.98}   & \bf{1.53}   & \bf{0.25}       \\ 
\bottomrule
\end{tabular}}
\label{table:main2}
\vspace{-5mm}
\end{table}

\subsection{Evaluation of Decoupling Ability}

\begin{figure}[htb]
\label{main_fig}
  \centering
  \includegraphics[width=0.9\linewidth]{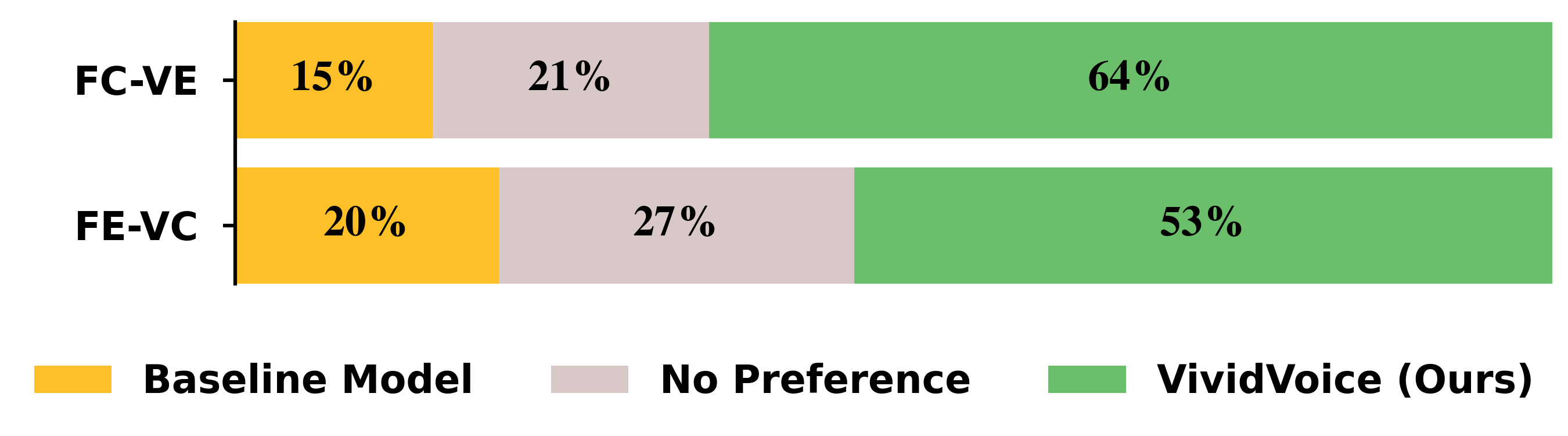}
  \vspace{-3mm}
  \caption{
A/B Preference Test Results for Decoupling Ability.} 
  \label{fig:ab}
  \vspace{-4mm}
\end{figure}

To explicitly evaluate the decoupling capability of VividVoice, we conducted an A/B preference test comparing our model with the baseline model (w/ Attn-Fusion). The test included two scenarios: Fixed Character, Varying Environment(FC-VE), and Fixed Environment, Varying Character(FE-VC). Evaluators were asked to choose which model better controlled the varying attribute independently while keeping the fixed attribute unchanged. As shown in Fig. \ref{fig:ab}, in the “Fixed Character” and “Fixed Environment” tests, VividVoice achieved a 64\% and 53\% preference rate, respectively. These results strongly demonstrate the significant superiority of our model in independently controlling timbre and environment.

\vspace{-2mm}
\section{Conclusion}

This paper introduces and defines a novel task—scene-aware visually driven speech synthesis—and presents the VividVoice framework as a unified solution for this task. To support this task, we systematically constructed a large-scale multimodal dataset, Vivid-210K. The core module, D-MSVA, with its unique decoupled memory bank design and cross-modal hybrid supervision strategy, successfully achieves fine-grained alignment from visual scenes to timbre and environmental acoustic features. Both subjective and objective experimental results provide strong evidence that VividVoice significantly outperforms existing models in terms of audio fidelity, content clarity, and multimodal consistency. Ablation experiments further validate the superiority of the D-MSVA module in improving audio quality compared to traditional fusion methods.

\bibliographystyle{IEEEbib}
\bibliography{refs}

\end{document}